\documentclass[aps,prl,twocolumn,superscriptaddress,nofootinbib,preprintnumbers]{revtex4}

\usepackage{graphicx}
\usepackage{graphics}
\usepackage{amsmath,amssymb}
\usepackage[usenames]{color}
\usepackage{subfigure}
\usepackage{hyperref}
\usepackage{breakurl}
\usepackage{enumitem}

\def\Tr{\,{\rm Tr}\,}

\newcommand{\da}{\dagger}

\newcommand{\be}{\begin{equation}}
\newcommand{\ee}{\end{equation}}
\newcommand{\bea}{\begin{eqnarray}}
\newcommand{\eea}{\end{eqnarray}}
\newcommand{\ben}{\begin{enumerate}}
\newcommand{\een}{\end{enumerate}}
\newcommand{\bit}{\begin{itemize}}
\newcommand{\eit}{\end{itemize}}

\newcommand{\la}[1]{\label{#1}}

\newcommand{\Eq}[1]{Eq.~(\ref{#1})}

\newcommand{\Fig}[1]{Fig.~\ref{#1}}

\def\a{\alpha}
\def\b{\beta}

\def\nl{\nonumber \\}

\newcommand{\vv}[1]{\mathbf #1}							% 3-vector
\newcommand\La{{\cal L}}

\newcommand{\bert}{\raise-0.45mm\hbox{\Large$\Box$}}			% D'Alembertian

\newcommand{\gd}{\gamma_\downarrow}						% Decay rate
\newcommand{\gu}{\gamma_\uparrow}						% Pumping rate

\makeatletter
\newcommand*\bigcdot{\mathpalette\bigcdot@{.5}}
\newcommand*\bigcdot@[2]{\mathbin{\vcenter{\hbox{\scalebox{#2}{$\m@th#1\bullet$}}}}}
\makeatother

\definecolor{BrickRed}{cmyk}{0,0.89,0.94,0.28}					%%%PANTONE 1805
\definecolor{MidnightBlue}{cmyk}{0.98,0.13,0,0.43}				%%%PANTONE 302
\definecolor{DarkGreen}{rgb}{0.100806,0.495968,0.209979}
\definecolor{orange}{rgb}{0.587167,0.354498,0.146197}

\begin{document}

\title{Quantum thermodynamics of coronal heating}

\author{Robert Alicki}\email{robert.alicki@ug.edu.pl}
\affiliation{International Centre for Theory of Quantum Technologies (ICTQT), University of Gda\'nsk, 80-308, Gda\'nsk, Poland}
\author{Alejandro Jenkins}\email{alejandro.jenkins@ucr.ac.cr}
\affiliation{International Centre for Theory of Quantum Technologies (ICTQT), University of Gda\'nsk, 80-308, Gda\'nsk, Poland}
\affiliation{Laboratorio de F\'isica Te\'orica y Computacional, Escuela de F\'isica, Universidad de Costa Rica, 11501-2060, San Jos\'e, Costa Rica}

\date{First version: 15 March 2021; this revision 21 February, 2022}

\begin{abstract}
Using the quantum Markovian master equation, we show that convection in the stellar photosphere generates plasma waves by an irreversible process akin to Zel'dovich superradiance and sonic booms.  In the Sun, this mechanism is most efficient in quiet regions with small magnetic fields.  Energy is mostly carried by megahertz Alfv\'en waves that scatter elastically until they reach a height at which they can dissipate via mode conversion.  This gives the right power flux for coronal heating and may account for chromospheric evaporation leading to impulsive heat transport in the corona.
\end{abstract}

\maketitle

%%%%%%%%%%
%%% INTRODUCTION
%%%%%%%%%%

{\it Introduction.}---Coronal heating is the transport of energy from the colder photosphere on a star's surface to the hotter corona in its upper atmosphere.  The second law of thermodynamics requires that this be driven by a macroscopic engine.  Convection cells (granules or supergranules) on the photosphere, powered by temperature and gravity gradients, can make Alfv\'en waves that propagate upwards in the stellar atmosphere and dissipate in the corona \cite{Alfven}.  This picture is supported by observations of Alfv\'en waves generated in the magnetic network of the Sun's photosphere, with typical frequencies in the millihertz scale.  Theorists have proposed mechanisms, based on flux-tube shaking, shocks, and turbulence, that generate low-frequency Alfv\'en waves traveling in open or closed magnetic tubes; see, e.g., \cite{Cranmer2005}.  These models have limitations, however, including difficulties in accounting for the thermalization of Alfv\'en waves in the corona.  For a recent review, see \cite{review2020}.

Some solar physicists have argued that magnetohydrodynamic (MHD) models and codes must be supplemented by ``nonclassical particle transport'' \cite{Klimchuk}.  In this letter we show that coronal heating may be explained if the production, propagation, and dissipation of Alfv\'en waves in the stellar atmosphere are treated in terms of quantum thermodynamics.  Steady convective circulation in the photosphere generates a non-equilibrium spectrum of Alfv\'en waves by an irreversible process similar to Zel'dovich's rotational superradiance \cite{Zeldovich1, Zeldovich2, Bekenstein, BCP} and the Ginzburg-Frank theory of radiation by uniformly moving sources \cite{Ginzburg-Frank, Ginzburg1, Ginzburg2} (including sonic booms and \u{C}erenkov radiation).  This extends and applies previous work by the authors on formulating superradiance and related processes in terms of quantum fields coupled to moving baths \cite{rotatingbath, tribo}.

For the Sun, most of the superradiant Alfv\'en waves come from quiet regions where the magnitude of the magnetic field is of order one gauss.  Energy is mostly carried by waves with megahertz-scale frequencies, much higher than the frequencies considered in previous wave heating models.  We will argue that such high-frequency modes diffuse upwards in the inhomogeneous medium, before thermalizing in the upper chromosphere via mode conversion.  Observation shows that nanoflares in active regions have enough power to account for much of the coronal heating \cite{flares}, but this raises the question of why repeated flares should result in net {\it upward} transport of energy, against the underlying temperature gradient.  We propose that such flares result from chromospheric evaporation \cite{paradox} powered by superradiant Alfv\'en waves heating the plasma from below.  Our model may thus transcend the dichotomy in solar physics between wave and impulsive heating models \cite{Antolin}.

%%%%%%%%%%
%%% QUANTUM VIEWPOINT
%%%%%%%%%%

{\it Quantum viewpoint.}---We apply the methods of quantum thermodynamics to the production and absorption of Alfv\'en waves by thermal plasmas (see Supplemental Material for more details on this approach and how it connects to the general theory of open quantum systems).  Consider a harmonic oscillator with mass $m=1$ and angular frequency $\omega$.  It can be described using position $x$ and momentum $p$ operators with canonical commutor $[x, p] = i\hbar$ or, more conveniently, in terms of complex amplitudes (annihilation and creation operators),
\be
a = \sqrt{\frac{1}{2 \omega \hbar}} (\omega x + i p)  ~~\hbox{and}~~ a^\da = \sqrt{\frac{1}{2 \omega \hbar}} (\omega x - i p) ,
\la{ec:aa}
\ee
so that $[a, a^\da] = 1$.  In terms of the number operator $n = a^\da a$, the Hamiltonian (energy operator) is
\be
H = \hbar\omega n ,
\la{eq:H}
\ee
ignoring the vacuum contribution $\hbar \omega / 2$.

If this oscillator is weakly coupled to a large bath in a stationary state, via an interaction Hamiltonian linear in $x, p$, and if the usual conditions for the validity of the Markovian approximation are fulfilled, the non-unitary evolution of the reduced density matrix $\rho (t)$ is described by the master equation
\bea
\frac{d \rho}{dt} &=& - i \omega [n , \rho ] + \frac 1 2 \gd \bigl([ a ,\rho  a^\da ] + [ a \rho , a^\da ] \bigr) \nl
&& + \frac 1 2 \gu \bigl([ a^\da ,\rho  a] +[ a^\da \rho , a] \bigr) ,
\la{eq:MME} 
\eea   
where $\gd, \gu$ are, respectively, the damping and pumping rates.  These can be expressed in terms of the Fourier transforms of reservoir autocorrelation functions for the relevant observables in the oscillator-bath interaction, computed in the bath's stationary state \cite{AL}.  Alternatively, the same expressions can be simply computed from Fermi's golden rule \cite{golden}.

If the bath is in equilibrium at temperature $T$, the damping and pumping rates satisfy the Kubo-Martin-Schwinger (KMS) condition
\be
\frac{\gu}{\gd} = e^{-\b \hbar\omega} ,
\la{eq:KMS}
\ee
where $\b = 1 / (k_B \, T)$ is the inverse temperature ($k_B$ being the Boltzmann constant).  Equation \eqref{eq:KMS} implies that any initial state $\rho(0)$ will thermalize:
\be
\lim_{t \to \infty} \rho(t) =  Z^{-1} e^{-\b H} ,    
\la{eq:thermal}
\ee
where $Z$ is the partition function.

The simplicity of \Eq{eq:MME} allows us to find exact solutions in various representations. Observables
\be
\a (t) = \Tr [\rho(t) a] \quad \hbox{and} \quad \bar n = \Tr [\rho (t) n]
\la{eq:observables}
\ee
obey closed evolution equations:
\bea
\frac{d \a}{dt} &=& \left[ -i \omega - \frac 1 2 (\gd -\gu) \right] \a , \la{eq:field} \\
\frac{d \bar n}{dt} &=& - (\gd -\gu) \bar n + \gu  =  - \gd  \bar n+ \gu (1+\bar n) . \la{eq:pnumber}
\eea
In quantum field theory, the oscillator can be interpreted as a single mode for waves in a cavity, so that Eqs.\ \eqref{eq:field} and \eqref{eq:pnumber} reflect wave-particle duality.

A classical wave picture based on \Eq{eq:field} gives a correct energy balance only in very special cases.  For a zero-temperature bath ($\gu = 0$) and an initially coherent state, the state remains coherent, i.e.,
\be
\rho(t) = |\a(t)\rangle\langle\a(t)| , ~\hbox{for}~ t \geq 0, 
\la{eq:coherent}
\ee
with $\a(t)$ satisfying \Eq{eq:field}.  The mode's energy
\be
E(t) = \Tr [\rho(t) H] = \hbar \omega \bar{n}(t)
\ee
takes the form
\be
 E(t) = \hbar \omega \langle \a(t) | a^\da a | \a(t) \rangle =  \hbar \omega | \a(t)|^2  .
\la{eq:cohenergy}
\ee
In this case the ``classical wave equation with damping'' (\Eq{eq:field}) completely describes the evolution of the fundamental measurable quantities: the field amplitude and the energy of the mode. This remains true if the oscillator is driven by an external deterministic and time-dependent force described by a Hamiltonian of the form
\be
H(t) =  \hbar\omega  a^\da a  + \bar \xi (t) a  + \xi(t) a^\da .
\la{eq:Ham}
\ee

Only in these very simple cases can the energy balance be obtained without recourse to ``particle'' numbers.  For a wave propagating in an inhomogeneous medium, elastic scattering causes strong decoherence, which can be described by adding a decoherence rate $\Gamma > 0$ to the quantity \hbox{$(\gd -\gu)$} in \Eq{eq:field} only \cite{waves}.  Evidently, the solution to the classical wave equation (\Eq{eq:field})
\be
\a(t) = e^{\frac 1 2 (\gu-\gd - \Gamma)t} \, e^{-i\omega t} \a(0) ,
\la{eq:fieldsol}
\ee
does not give the right energy balance, since the decoherence rate $\Gamma$ may prevent $| \a (t) |^2$ from growing despite active pumping $\gu > \gd$.

The case $\gu > \gd$ corresponds to non-equilibrium stationary reservoirs, like a macroscopically moving heat bath or an optically active medium with population-inverted atomic levels.  The resulting dynamics is known in black-hole thermodynamics as ``superradiance'' \cite{BCP} and in quantum optics as ``laser action'' \cite{Carmichael}.  It may be described in terms of {\it negative-temperature} baths, as suggested by \Eq{eq:KMS}.  We can obtain the correct balance from \Eq{eq:pnumber}, which implies exponential energy growth
\be
E(t) = E(0) \, e^{(\gu-\gd) t}   + \hbar \omega \frac{\gu}{\gu-\gd} \left[ e^{(\gu-\gd)t} -1 \right] .
\la{eq:energy}
\ee
Such growth results from {\it stimulated emission}, a quantum phenomenon described by the term in \Eq{eq:pnumber} proportional to $\bar n$.

%%%%%%%%%%
%%% ANOMALOUS DOPPLER SHIFT
%%%%%%%%%%

{\it Anomalous Doppler shift.}---Consider a single mode of a quantum field in a cavity, characterized by wave vector $\vv k$ and angular frequency $\omega(\vv k)$.  If this mode is coupled to a bath moving with respect to the cavity at velocity $\vv v$, the mode's frequency is Doppler-shifted in the reservoir's frame of reference,
\be
\omega \to \omega'(\vv k) = \omega(\vv k) -  \vv k \cdot \vv v ,
\la{eq:Doppler}
\ee
and the KMS condition of \Eq{eq:KMS} becomes 
\be
\frac{\gu (\vv k)}{\gd (\vv k)} =  e^{\b \hbar [\vv k \cdot \vv v - \omega(\vv k)]} =e^{-\b_{\rm loc} (\vv k ) \hbar \omega(\vv k)} ,
\la{eq:newKMS}
\ee
where the ``local'' inverse temperature is
\be
\b_{\rm loc} (\vv k ) = \b \cdot \left[1 - \frac{\vv k \cdot \vv v} {\omega(\vv k)}\right] .
\la{eq:negbeta}
\ee
Thus, for modes satisfying
\be
\omega(\vv k) \leq \vv k \cdot \vv v ,
\la{eq:critical}
\ee
the moving reservoir acts as a negative-temperature bath, at the inverse temperature $\b_{\rm loc} (\vv k ) < 0$, so that it can amplify the mode's energy.  In their theory of radiation by uniformly moving sources, Frank and Ginzburg called this an ``anomalous Doppler shift'' \cite{Ginzburg-Frank}.  See Supplemental Material for details on how this connects to Zel'dovich's theory of rotational superradiance.

%%%%%%%%%%
%%% ALFVEN-WAVE SUPERRADIANCE
%%%%%%%%%%

{\it Alfv\'en-wave superradiance.}---In the two-fluid theory of partially ionized hydrogen plasma \cite{2fluid}, the Alfv\'en-wave speed is given by
\bea
\hskip -12 pt v_A &=& \frac{B}{\sqrt{\mu_0 \rho}} \nl
&=&  \left( 2.18 \times 10^{11} ~ \frac{\rm cm}{\rm s} \right) \left( \frac{N}{{\rm cm}^{-3}} \right)^{-1/2} \left( \frac{B}{\rm gauss} \right)
\la{eq:vA}
\eea
in two extreme cases:
\begin{enumerate}[label=(\alph*)]
	\item for wave frequencies much lower than the ion-neutral collision frequency, with $N= N_H$ (density of hydrogen atoms), or
	\item for wave frequencies much higher than the ion-neutral collision frequency, with $N= N_I$ (density of hydrogen ions).
\end{enumerate}
Consider a quiet region of the Sun's photosphere, with magnetic field of order one gauss, a regime that makes up most of the Sun's surface (see, e.g., \cite{internetwork}).  Then \hbox{$N_H = 1.2 \times 10^{17} ~{\rm cm}^{-3}$} and \hbox{$N_I = 6.4 \times 10^{13} ~{\rm cm}^{-3}$}, while the ion-neutral collision frequency is \hbox{$\nu_{\rm in} = 1.2 \times 10^9$ Hz} \cite{Pandey}.  As we shall corroborate later, the cutoff in Alfv\'en-wave frequency allows us to restrict our attention to case (a).  For $B \simeq 1$ G, this gives \hbox{$v_A \simeq 6$ m/s}.  Taking the sound speed at the surface as \hbox{$v_s \simeq 10$ km/s} and the average speed of granular flow as \hbox{$v \simeq 1$ km/s} \cite{convection}, we find that
\be
v_A \ll v < v_s \ll c .
\la{eq:vels}
\ee
Note that within the flux tubes of the photospheric network, with kilogauss magnetic fields, the first inequality is not satisfied.

\begin{figure} [t] 
\begin{center}
	\includegraphics[width=0.32 \textwidth]{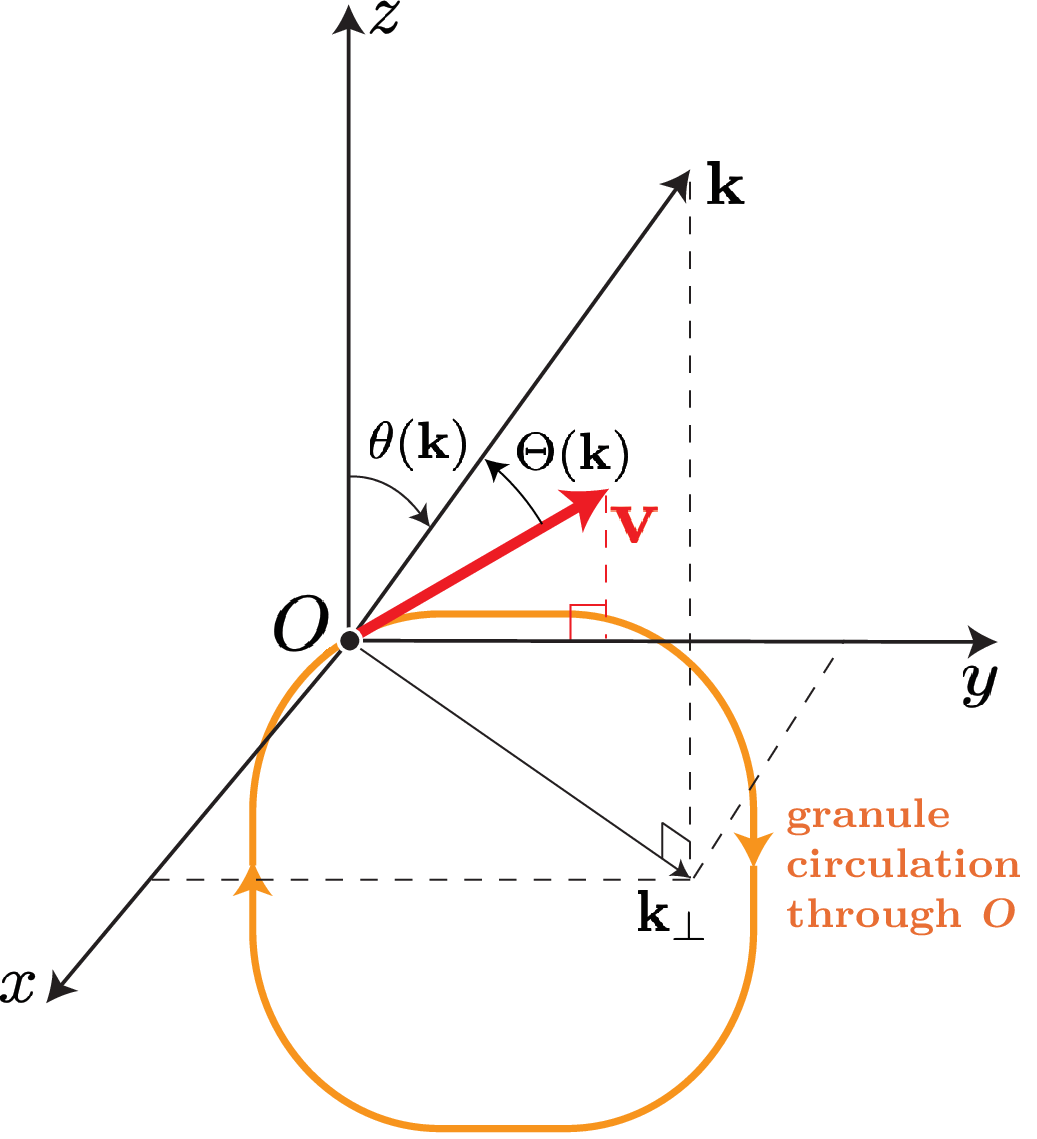}
\end{center}
\caption{\small Geometric setup for calculating Alfv\'en-wave superradiance: $\vv k$ is the wave vector of the relevant Alfv\'en mode, $\vv v$ the flow velocity corresponding to the granule circulation through a point $O$ on the photosphere, $\theta (\vv k)$ the angle from the $z$ axis to $\vv k$, and $\Theta (\vv k)$ the angle from $\vv v$ to $\vv k$.  The magnetic field points along the $z$ axis (i.e., vertically).  The projection of $\vv k$ onto the horizontal $x$-$y$ plane is labelled $\vv k_\perp$.  For clarity, $\vv v$ is drawn on the $y$-$z$ plane, but this is not assumed in the analysis.\la{fig:geometry}}
\end{figure}

The inequalities in \Eq{eq:vels} imply that pure Alfv\'en waves and slow waves propagate with approximately the same phase velocity
\be
v_A (\vv k) = v_A \left| \cos \theta(\vv k) \right| ,
\la{eq:vA1}
\ee
where $\theta(\vv k)$ is the angle between the wave vector $\vv k$ and the vertical magnetic field (see \Fig{fig:geometry}).  The remaining MHD waves propagate with a phase velocity approximately isotropic and equal to $v_s$.  Because the flow speed $v$ is smaller than $v_s$, \Eq{eq:critical} implies that only pure Alfv\'en and slow waves can be superradiated.  From now on we will refer to both as ``Alfv\'en waves'' and to the corresponding quasiparticles as {\it alfvenons}. 

High-frequency Alfv\'en waves generated at a square patch of dimensions $\ell \times \ell$ on the stellar surface can be decomposed into plane waves propagating with the phase velocity of \Eq{eq:vA1}.  Because the upper boundary is the stellar corona, which can be treated as a reservoir at an effectively infinite temperature (compared to the photosphere's $T$), only superradiant modes (for which $\b_{\rm loc} (\vv k) < 0$ in \Eq{eq:negbeta}) can transport energy towards the corona.  Our results will not depend on $\ell$, as long as $v_A$, $v$, and $v_s$ can be taken as constant within the patch.

The dispersion relation for Alfv\'en waves is
\be
\omega(\vv k) = v_A |k_z| = v_A k \left| \cos \theta(\vv k) \right| , \quad  k = |\vv k| .
\la{eq:dispersion}
\ee
The condition of \Eq{eq:critical} then becomes
\be
\Theta(\vv k) \in [0, \pi / 2] , \quad \cos\Theta(\vv k) \geq \frac{v_A}{v} \left| \cos \theta(\vv k) \right|,   
\la{eq:critical2}
\ee
where $\Theta (\vv k)$ is an angle between $\vv k$ and $\vv v$ (see \Fig{fig:geometry}).  If \hbox{$v_A / v \ll 1$}, then \Eq{eq:critical2} implies that practically all Alfv\'en modes with $\cos\Theta(\vv k) \geq 0 $ and  $\cos\theta(\vv k)\geq 0$ contribute to an irreversible upwards transport of energy, driven by superradiance.

The maximal frequency $\Omega_A$ of the Alfv\'en modes is bounded by $k_A v_A$, where $k_A$ is the maximal wave-vector magnitude, which we can roughly estimate from the inverse of the typical distance between neighboring ions.  Thus
\be
k_A \simeq N_I^{1/3} \quad \hbox{and} \quad \Omega_A \simeq  N_I^{1/3}v_A ,
\la{eq:cutoff}
\ee
where $N_I$ is the number of ions per unit volume (see Supplemental Material for further discussion of this cutoff).  For the Sun's photosphere (\hbox{$N_I \simeq 10^{20}~{\rm m}^{-3}$}; \hbox{$v_A \simeq 6$ m/s}) this gives \hbox{$\Omega_A \simeq 10^7$ Hz}.  For temperature \hbox{$T \simeq 6,000$ K} we have
\be
\frac{\hbar \Omega_A}{k_B T} \simeq 3 \times 10^{-8} \ll 1,
\la{eq:ratio}
\ee
which implies that, in equilibrium, each mode carries energy $k_B T$ (equipartition).  Since $\Omega_A$ is well below the ion-neutral collision frequency at the photosphere, taking $N = N_H$ in \Eq{eq:vA} was justified.

%%%%%%%%%%
%%% POWER FLUX
%%%%%%%%%%

{\it Power flux.}---The temperature above the Sun's surface and below the corona is approximately constant and close to the surface $T$ \cite{Erdelyi}.  Statistical alfvenon occupation numbers are therefore approximately equal to their equilibrium values and hence, taking into account \Eq{eq:ratio},
\be
\bar n_{\rm stat}(\vv k) = \frac{k_B T}{\hbar \omega(\vv k)} .
\la{eq:nstat}
\ee
Additional non-thermal alfvenons produced superradiantly are transported upwards towards the corona.  Under temporally stationary conditions,
\be
0 = -[ \gd(\vv k) -\gu(\vv k)] \bar n_{\rm stat}(\vv k) +    \gu(\vv k) - \Gamma_{\rm diss}(\vv k) ,
\la{eq:balance}
\ee
where the term $\Gamma_{\rm diss}(\vv k)$ accounts for the power deposited in the corona.

Inserting Eqs.\ \eqref{eq:newKMS} and \eqref{eq:nstat} into \Eq{eq:balance}, and using \Eq{eq:ratio} to simplify the result, we find that the steady power carried away by the superradiant Alfv\'en modes from a patch of unit area at the surface is
\bea
J_A &=& \frac{1}{\ell^2} \sum_{\{+,+\}} \Gamma_{\rm diss}(\vv k) \hbar \omega(\vv k) = \frac{1}{\ell^2} \sum_{\{+,+\}} \gu(\vv k) \hbar \vv k \cdot \vv v \nl
&=& \frac{v}{\ell^2} \sum_{\{+,+\}} \gu(\vv k) \hbar k \cos \Theta(\vv k) \nl
&=& \langle \cos \Theta \rangle \frac{v}{\ell^2} \sum_{\{+,+\}}\gu(\vv k) \hbar k ,
\la{eq:flow}
\eea
where the sum over $\{+,+\}$ corresponds to wave vectors such that $\cos \theta(\vv k)$ and $\cos \Theta(\vv k)$ are both positive.

We may find $J_A$ without explicitly computing the $\gu (\vv k)$'s, by comparing \Eq{eq:flow} to the expression for thermal radiation.  Let $J_{\rm eq}(T)$ be the power flux carried by alfvenons emitted (or absorbed) in equilibrium at temperature $T$:
\bea
J_{\rm eq} (T) &=& \frac{1}{\ell^2}\sum_{\{+\}} \gu(\vv k) \hbar \omega(\vv k) = \frac{v_A}{\ell^2}\sum_{\{+\}} \gu(\vv k) \hbar k \cos \theta(\vv k) \nl
&=& \langle \cos \theta \rangle \frac{v_A}{\ell^2}\sum_{\{+\}} \gu(\vv k) \hbar k \nl
&=& 2 \langle \cos^2\theta\rangle v_A  N_I k_B T ,
\la{eq:emit}
\eea
where the sum over $\{+\}$ corresponds to wave vectors such that $\cos \theta(\vv k)$ is positive (upward direction).  In the last equality we used a simple model of a ``black-body'' surface radiating alfvenons upwards with the $z$-component of the velocity equal to \hbox{$v_A^z = v_A \cos \theta$} (see \Eq{eq:dispersion}) and in the high-temperature regime (see \Eq{eq:ratio}).  Taking into account that \hbox{$\sum_{\{+\}} \cdots = 2 \sum_{\{+,+\}} \cdots$} and comparing Eqs.\ \eqref{eq:flow} and \eqref{eq:emit}, we conclude that
\be
J_A = \kappa  v N_I k_B T .
\la{eq:flow1}
\ee
The geometric factor $\kappa$ in \Eq{eq:flow1} is given by
\be
\kappa = \frac{\langle \cos \Theta \rangle \langle \cos^2 \theta \rangle}{\langle \cos \theta \rangle} 
\la{eq:flow2}
\ee
and is bounded as $1/3 < \kappa < 1$, with the lower bound corresponding to uncorrelated directions of $\vv k$ and $\vv v$.

Note that \Eq{eq:flow1} does not depend on the local $v_A$, making it insensitive to the large variations in the magnetic field, as long as the inequality of \Eq{eq:vels} holds.  For the Sun's atmosphere
\be
J_A \sim 10^4 ~{\rm W}/ {\rm m}^2 ,
\la{eq:JA}
\ee
consistent with the $10^3 - 10^4 ~{\rm W} / {\rm m}^2$ needed to account for coronal heating \cite{review2020}.

Since mode density scales as $\sim k^2$ and equipartition of energy holds, energy is mostly transported by the waves with megahertz frequencies, with micrometer-scale wavelengths at the surface.  Low frequency cutoffs due to various mechanisms proposed, e.g., in \cite{2fluid, cutoffs}, concern frequencies small compared to $\Omega_A$ and can therefore be neglected in the estimate of \Eq{eq:JA}.

%%%%%%%%%%
%%% SCATTERING AND DISSIPATION
%%%%%%%%%%

{\it Scattering and dissipation.}---Solar physicists have questioned the role of Alfv\'en waves in coronal heating because they appear difficult to dissipate.  This is true for low-frequency (millihertz scale) modes propagating in magnetic flux tubes with large (kilogauss scale) magnetic fields at the surface.  However, for the megahertz-scale frequencies relevant to our model, decoherence and dissipation processes for Alfv\'en waves must be re-examined.  High-frequency superradiant modes should exhibit much stronger elastic scattering in a nonuniform medium, since such processes typically scale with a positive power of the frequency. This gives strong decoherence {\it without} dissipation \cite{waves}.

``Mode conversion'' (see, e.g., \cite{dissipation}) also needs to be reconsidered for megahertz alfvenons.  Mode conversion is a transition between an alfvenon and a phonon, satisfying energy-momentum conservation
\be
E = \hbar \omega_1 = \hbar \omega_2 ; \quad \vv p = \hbar \vv k_1 = \hbar \vv k_2 .
\ee
This kinematic constraint can be fulfilled only if the local phase speed (\hbox{$v_{\rm ph} = \omega / | \vv k |$}) is the same for Alfv\'en and magneto-acoustic waves.  Due to momentum conservation, alfvenon-phonon conversion is collinear and can be analyzed one-dimensionally. The conversion rate, given by Fermi's golden rule, is proportional to frequency \cite{DiracEq}.  It is therefore enhanced by nine orders of magnitude if megahertz rather than millihertz modes are considered.  Once converted to phonons, the superradiant energy dissipates quickly.

\begin{figure} [t] 
\begin{center}
	\includegraphics[width=0.32 \textwidth]{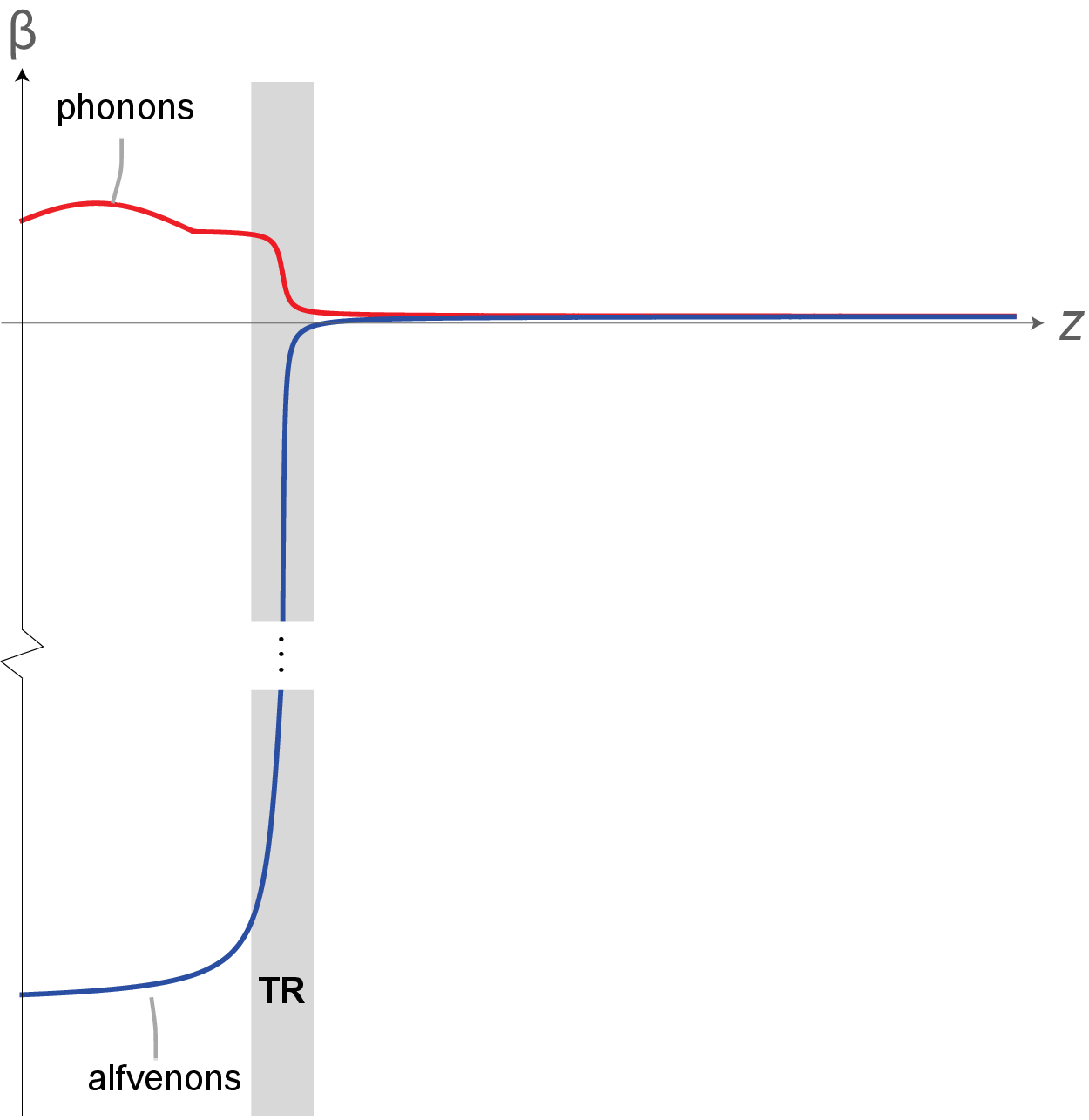}
\end{center}
\caption{\small Inverse temperature $\beta$ for phonons (red) and superradiant alfvenons (blue), as functions of height $z$ in the solar atmosphere.  Below the transition region (TR) the absolute value of $\beta$ is two orders of magnitude greater for alfvenons than for phonons [see \Eq{eq:negbeta}].\la{fig:beta-z}}
\end{figure}

In quiet regions of the surface of the Sun's photosphere
\be
v_A \simeq 6 ~\hbox{m/s} \ll v_s  \simeq 10 ~\hbox{km/s} ,
\ee
while in the corona
\be
v_A \simeq 10^3 ~\hbox{km/s} > v_s \simeq 10^2 ~\hbox{km/s} 
\ee
(see, e.g., \cite{pluto}).  The two speeds must match at an intermediate height $z$.  We propose that this occurs in the upper chromosphere, just below the ``transition region'' (TR) between chromosphere and corona.  This may explain why the observed (i.e., phonon) temperature rises so abruptly in the TR, at $z \simeq $ 2,100 km \cite{Erdelyi, pluto}, since the TR would be just above where superradiant alfvenons start mode-converting with phonons.  As illustrated by \Fig{fig:beta-z}, the negative temperature of superradiant alfvenons makes our model consistent with the basic thermodynamic principle that energy flows irreversibly towards increasing $\beta$.  (See Supplemental Material for further discussion of energy transport and dissipation in the stellar atmosphere.)

%%%%%%%%%%
%%% DISCUSSION
%%%%%%%%%%

{\it Discussion.}---Our model is an extension to MHD waves of previous work on superradiance from the perspective of quantum thermodynamics \cite{rotatingbath,tribo}.  Even though we considered hot plasmas (\hbox{$k_B T \gg \hbar \omega$}) and the final results were independent of $\hbar$, this treatment did not just clarify the relevant microphysics, but also simplified computing macroscopic observables.

Superradiant production of Alfv\'en waves by steady granular convection should be incorporated in models of stellar atmospheres.  Applied to the Sun, this gives an estimate, depending only on directly measurable parameters, of the power flux carried by superradiant Alfv\'en waves from the quiet regions of the solar photosphere to the corona [see \Eq{eq:flow1}].  Most of this power is carried by the shortest (micrometer scale) wavelengths.  Given the magnitude of the power flux and the fact that modes with such short wavelengths can dissipate via mode conversion, we find that Alfv\'en-wave superradiance can plausibly account for most of the heating of the Sun's corona.  Since superradiant waves can ``boil'' the plasma in the upper chromosphere \cite{paradox}, impulsive heating (by nanoflares) and wave heating (by Alfv\'en waves) may be aspects of one underlying mechanism. \\

%%%%%%%%%%
%%% ACKNOWLEDGEMENTS
%%%%%%%%%%

\begin{acknowledgments} We thank Zdzis{\l}aw Musielak for feedback and constructive criticism.  RA was supported by the Foundation for Polish Science's International Research Agendas, with structural funds from the European Union (EU) for the ICTQT.  AJ was supported by the Polish National Agency for Academic Exchange (NAWA)'s Ulam Programme (project PPN/ULM/2019/1/00284, ``Energy conversion by open quantum systems: Theory and applications''). \end{acknowledgments}

%%%%%%%%%%
%%% SUPPLEMENTAL MATERIAL
%%%%%%%%%%

\bibliographystyle{h-physrev}
\bibliography{ref}
\clearpage
\newcommand{\beginsupplement}{%
        \setcounter{table}{0}
        \renewcommand{\thetable}{S\arabic{table}}%
        \setcounter{figure}{0}
        \renewcommand{\thefigure}{S\arabic{figure}}%
        \setcounter{equation}{0}
        \renewcommand{\theequation}{S\arabic{equation}}%
     }
\onecolumngrid
\beginsupplement

\section{Supplemental Material}

To help clarify the motivation and the nature of our approach to the problem of coronal heating, let us briefly review the main results in the theory of open quantum systems that are relevant to this work.  The references that we provide here are to modern reviews that present the material from a perspective similar to our own, not to original sources.

Since we argue in our Letter that the macroscopic coronal heating problem is best understood in terms of Alfv\'en-wave quasiparticles ({\it alfvenons}) being produced or absorbed by thermal plasmas, let us begin by noting that the first evidence of the failure of classical physics was the falsification of the predictions of the equipartition theorem for the heat capacities of gases and solids \cite{FL}.  This was followed some decades later by the ultraviolet catastrophe in the classical theory of black-body radiation \cite{Weinberg}.

Irreversible processes involving the generation and absorption of waves by thermal environments may not always be well described by classical physics, even for hot and macroscopic media \cite{waves}.  One key reason for this is that, according to statistical physics, the generation of entropy by irreversible processes needs to be understood in terms of a counting of accessible quantum microstates.  Thus, Parikh, Wilczek, and Zahariade have recently considered how the statistical properties of thermal noise may serve as evidence of the quantization of gravitational waves \cite{grav1, grav2}.

Moreover, it is well known that the classical hydrodynamic description of shock waves leads to mathematically singular fronts \cite{classical}.  The application of the Markovian master equation (MME) to quantum field theory and to the statistical physics of waves is therefore a broadly important subject, but most of the work published so far has focused exclusively on quantum optics, especially laser physics \cite{Carmichael}.

\subsection{Open-system formalism}
\la{sec:open}

The MME for the quantum harmonic oscillator weakly coupled to a thermal bath [\Eq{eq:MME} in the Letter] has a long history that stretches back to Landau's early work in 1927.  Similar results were independently re-discovered and developed several times during the 20th century and in various contexts \cite{historyGKLS}.  In recent decades the irreversible dynamics of work extraction by a quantum system coupled to an external disequilibrium (to which the MME can be applied in a weak-coupling limit) has become a subject of much theoretical and practical interest in quantum thermodynamics \cite{QT}.  Note that the Markovian approximation fails if correlations within the bath do not decay quickly enough, implying that the bath cannot be characterized by one temperature and one chemical potential \cite{AL}.

The starting point of such a theory is the quantum MME for the reduced density matrix $\rho(t)$ of the open system of interest:
\be
\frac{d}{dt} \rho(t) = -\frac i \hbar \left[ H(t) , \rho(t) \right] + \La(t) \rho(t) ,
\la{eq:MME_s} 
\ee   
where the generator $\La(t)$ (also called ``superoperator'' or ``dissipator'') possesses a canonical structure, known as the Lindblad or GKLS form \cite{AL}. The time-dependence of the system Hamiltonian $H(t)$ and the generator $\La(t)$ can incorporate varying external conditions. The mathematical properties of \Eq{eq:MME_s} allows us to derive the first and second law of thermodynamics for such systems.  The multi-mode generalization of Eq.\ (3) gives: 
\be
\frac{d \rho}{dt} = - i \sum_{\vv k} \omega(\vv k) [n_{\vv k} , \rho ] + \frac 1 2 \sum_{\vv k} \gd(\vv k) \left( \left[ a_{\vv k}, \rho  a_{\vv k}^\da \right] + \left[ a_{\vv k} \rho , a_{\vv k}^\da \right] \right) + \frac 1 2 \sum_{\vv k} \gu(\vv k) \left( \left[ a_{\vv k}^\da, \rho  a_{\vv k} \right] + \left[ a_{\vv k}^\da \rho , a_{\vv k} \right] \right) .
\la{eq:MME_m} 
\ee
By a straightforward extension of the results derived by the authors in \cite{rotatingbath}, the dynamics of a quantum field coupled to a heat bath moving macroscopically with velocity $\vv v$ can be described in terms of \Eq{eq:MME_m} and the modified KMS relation  
\be
\frac{\gu(\vv k)}{\gd(\vv k)} =  e^{\b \hbar [\vv k \cdot \vv v - \omega(\vv k)]} .
\la{eq:KMS_s}
\ee

\subsection{Thermodynamics of quantum field}
\la{sec:QTF}

Introducing the averaged mode occupation numbers $\bar n_{\vv k}\equiv \Tr{(\rho n_{\vv k})} $, with the corresponding time derivative denoted by $\dot{\bar n}_{\vv k}$ and the internal energy of the the field by 
\be
U \equiv \Tr{(\rho H)} = \hbar \sum_{\vv k} \omega(\vv k) \bar n_{\vv k} ,
\la{eq:intenergy}
\ee
we can derive the first law of thermodynamics in the form
\be
\frac{dU}{dt} = J  + P 
\la{eq:Ilaw}
\ee
where 
\be
J  = \hbar \sum_{\vv k} \left[ \omega(\vv k) - \vv k \cdot \vv v \right] \dot{\bar n}_{\vv k}
\la{eq:heatcurrent}
\ee
is the net heat current supplied by the heat bath to the field, and 
\be
P  = \vv v \cdot \sum_{\vv k} \hbar\vv k \cdot \dot{\bar n}_{\vv k} 
\la{eq:power}
\ee
is the mechanical power supplied by the macroscopic motion of the bath.  The second law of thermodynamics is then obtained in the form
\be
\frac{dS}{dt} - \frac J T \ge 0,
\la{eq:IIlaw}
\ee
where the entropy $S(t)$ is identified with the quantum von Neumann entropy of the reduced density matrix
\be
S = - k_B \Tr{(\rho \ln \rho)} .
\la{eq:entropy}
\ee
This approach gives us a full quantum-thermodynamic treatment of the interaction between the field and the bath, provided that the conditions for the validity of the Markovian approximation in \Eq{eq:MME_s} are met \cite{AL}.

\subsection{High-frequency cutoff for Alfv\'en waves}
\la{sec:Debye}

Our estimate of the high-frequency cutoff $\Omega_A$ for the Alfv\'en waves [\Eq{eq:cutoff} in the Letter] appears similar to the Debye frequency for a solid \cite{Debye}.  We expect that the number of relevant degrees of freedom per unit volume should be $2 N_I$, rather than $3 N_I$, since only the motion of the ions perpendicular to the magnetic field can make Alfv\'en waves.  The stellar plasma is more akin to a liquid than to a solid, but it has recently been established that the Debye model can be usefully applied to liquids, especially at the high frequencies that dominate the liquid's thermodynamic properties \cite{liquid}.  Since, for the present purposes, we need only an order-of-magnitude estimate of $\Omega_A$ at the photosphere (where the superradiant Alfv\'en waves are generated), we will not attempt a more detailed modelling of the physics that cuts off the high-frequency MHD modes.

The MHD formalism is non-relativistic, since it neglects the displacement currents in the Maxwell equations for the electromagnetic fields.  This is referred to as the ``quasi-stationary'' regime \cite{MHD}.  Since we estimate the cutoff frequency as \hbox{$\Omega_A \simeq 10^7$ Hz}, and the typical distance between neighboring ions at the photosphere is
\be
d \simeq N_I^{-1/3} \simeq 10^{-7} ~\hbox{m} ,
\la{eq:ion-d}
\ee
the speed of the motion of the ions does indeed remain highly non-relativistic all the way up to the cutoff:
\be
d \cdot \Omega_A \simeq 1 ~\hbox{m/s} \ll c .
\la{eq:non-rel}
\ee

\subsection{Superradiant energy transport and dissipation}
\la{sec:superrad}

In 1971, Zel'dovich argued that a rotating dielectric's kinetic energy could be partially converted to coherent radiation \cite{Zeldovich1, Zeldovich2}.  This effect, now called {\it superradiance}, results from the anomalous Doppler shift when the dielectic moves faster than the phase velocity of an incident radiation mode.\footnote{This is qualitatively different from the equilibrium phenomenon, also called ``superradiance'', first described by Dicke in \cite{Dicke}.}  For modes that fulfill the ``anomalous Doppler shift'' condition [\Eq{eq:critical} in the Letter], work may be extracted via stimulated emission, while entropy is produced in the rotating dielectric acting as a moving heat bath \cite{rotatingbath}.

Zel'dovich's prediction of rotational superradiance, which was independently of the research connected with the MME, played a key role in the development of black-hole thermodynamics and it provides a useful guide to a broad class of active, irreversible processes \cite{Bekenstein, BCP}.  The close analogy between Zel'dovich's theory of superradiance and the Ginzburg-Frank theory of radiation by uniformly moving sources (which includes \u{C}erenkov radiation, sonic booms, and other hydrodynamic shockwaves) was beautifully elucidated in \cite{Bekenstein}.

On the other hand, the fact that this same superradiance, considered from the point of view of the theory of open quantum systems, appears as a form of laser action was first clearly expressed by the authors in \cite{rotatingbath}.  That result led to an application of the MME formalism to fermion fields that offered a qualitatively new theory of triboelectricity in materials science \cite{tribo}.  In this picture, the ``anomalous Doppler shift'' condition [\Eq{eq:critical} in the Letter] can be interpreted in terms of {\it population inversion} [$\gu(\vv k) > \gd(\vv k)$ in \Eq{eq:KMS_s}] and a negative local temperature \cite{localT}, given by \Eq{eq:negbeta} in the Letter.  Since a negative temperature is hotter than any positive temperature, this helps to clarify how energy can flow persistently from the solar photosphere to the corona in a way consistent with statistical physics [see \Fig{fig:beta-z} in the Letter].

In the present work the treatment based on the MME has been applied, for the first time, to plasma physics.  This has allowed us to identify the Alfv\'en-wave analog of a sonic boom, providing a mechanism for energy transport not contemplated in any of the current MHD formalisms or codes used in stellar physics.  The quantum-thermodynamic approach also made it very straightforward to estimate the corresponding power flux [\Eq{eq:flow1} in the Letter].  In the Letter we have shown how this can resolve the long-standing problem of coronal heating for the Sun.  More than just showing ``how quantum mechanics helps us understand classical mechanics'' \cite{Zeldovich1, Paradoksov}, this illustrates why a quantum treatment of active transport processes in systems far from thermal equilibrium is needed in statistical physics.

Since the vertical magnetic fields in quiet regions of the solar atmosphere are anisotropic and distorted by plasma motions, we do not expect superradiant alfvenons to propagate straight towards the corona.  The dispersion relation [Eqs.\ \eqref{eq:vA} and \eqref{eq:vA1} in the Letter] implies that Alfv\'en waves can refract and reflect as they travel upwards along magnetic field lines.  Elastic scattering in the inhomogenous medium causes alfvenon decoherence without dissipation \cite{waves}.  We expect that the high-frequency (megahertz) alfvenons that carry most of the superradiant energy {\it diffuse} upwards in a process somewhat akin to the photon diffusion that transports energy from the stellar core to the photosphere.  When these superradiant alfvenons reach the upper chromosphere they begin to dissipate via mode conversion into phonons, as described in the Letter.

Although our model predicts that the superradiant alfvenons that provide the energy for coronal heating are mostly produced in quiet regions of the Sun's photosphere, this is not necessarily at odds with the expectation that much (or most) of the coronal heating occurs in active regions characterized by large magnetic fields.  Determining the consequences of our model for the impulsive dynamics of the transition region (TR) will require new and detailed numerical simulations.  Note that the increase in the Alfven velocity with the intensity of the magnetic field can allow scattered alfvenons to dissipate sooner in active as opposed to quiet regions of the upper chromosphere.  Moreover, some of the events associated with impulsive heating may result from loops in active regions being filled with surrounding plasma that has been heated from below \cite{paradox}.

\subsection{Parametric down-conversion}
\la{sec:PDC}

Nonlinear processes, usually neglected for low-frequency Alfv\'en waves, also become important at the high frequencies that carry most of the superradiant energy.  For instance, down-conversion of an alfvenon with wave vector $\vv k$ into two alfvenons with wave vectors $\vv k'$ and $\vv k''$ satisfying momentum conservation (\hbox{$\vv k = \vv k' + \vv k''$}, with equal signs for $k_z, k_z'$ and $k_z''$) automatically satisfies energy conservation as well, due to the particular form of the dispersion relation [\Eq{eq:dispersion} in the Letter]. The probability of this kinematically allowed process is proportional to the density of final states, which grows linearly with $| \vv k|$ and hence must be significant for high-frequency alfvenons, contributing to their dissipation.
\end{document}